\documentclass[11pt]{iopart}

\expandafter\let\csname equation*\endcsname\relax
\expandafter\let\csname endequation*\endcsname\relax 

\usepackage[english]{babel}
\usepackage[utf8]{inputenc}
\usepackage{amsmath}
\usepackage{graphicx}
\usepackage{subfigure}
\usepackage[colorinlistoftodos]{todonotes}
\usepackage{hyperref}
\usepackage{epsfig}
\usepackage{booktabs}
\usepackage{sidecap}

\begin{document}

\title{Spontaneous Segregation of Agents Across Double Auction Markets}

\author{Aleksandra Alori\'c$^\dag$, Peter Sollich$^\dag$, Peter McBurney$^\ddag$}
\address{$\dag$ ~ Department of Mathematics, King's College London, Strand,
London WC2R 2LS, UK}
\address{$\ddag$ ~ Department of Informatics, King's College London, Strand,
London WC2R 2LS, UK}

\ead{aleksandra.aloric@gmail.com}

\begin{abstract}
In this paper we investigate the possibility of spontaneous segregation into groups of traders that have to choose among several markets. Even in the simplest case of two markets and Zero Intelligence traders, we are able to observe segregation effects below a critical value $T_c$ of the temperature $T$; the latter regulates how strongly traders bias their decisions towards choices with large accumulated scores. It is notable that segregation occurs even though the traders are statistically homogeneous. Traders can in principle change their loyalty to a market, but the relevant persistence times become long below $T_c$.
\end{abstract}

\section{Introduction}
\label{Introduction}
Adam Smith, in his \textit{The Wealth of Nations} said that the concept of economic growth is deeply rooted in the division of labour. This primarily relates to the specialization of the labour force, where narrowing expertise allows better exploitation. Contemporary examples of such specialization include, e.g., airline companies: some specialize in first class and business flights, while others provide mainly low cost flights. The paper \cite{philippines} reports segmentation phenomena in the informal credit market in the Philippines, where lenders who specialize in trading make loans mainly to large and asset-rich farmers, while others lend more to small farmers and landless labourers. 

It can be argued that the space of customers is already segmented, and that the role of an efficient merchant is to find and adapt to niches in this customer space (see for example \cite{CATcosegregation}). However, here we want to explore the possibility of spontaneous segregation of initially {\em homogeneous} traders. This work was motivated by observations from \textit{the CAT Market Design Tournament} \cite{CAToverview} where competitors were invited to submit market mechanisms for a population of traders provided by the tournament organizers. It was observed that by co-adaptation of markets and traders the system evolved to a segregated state signalled by persistent ``loyalty'' of certain groups of traders to certain markets.

In order to test our hypothesis that segregation can emerge spontaneously, we have constructed a simple model of markets and traders. Markets are governed here by simple static sets of rules --- how to set the trading price and how to match traders. Traders are taken as Zero Intelligence agents following Ref.~\cite{ZItraders}. Such traders act largely randomly. This makes them a convenient tool for investigating the impact of market mechanisms \cite{ZILadley}, by removing all of the complexity associated with the traders' strategies. However, we note that our agents are zero intelligence only with respect to price, i.e.,\ they generate bids and asks at random. On the other hand they do learn from past successes or failures about the choice of market and whether to buy or sell.

Closely related work on segregation in \cite{luckystar} studies agents competing for parking spots in a one way street. A learning process is again used, with rewards (the closer to the city centre the better) and penalties (an agent is punished if s/he reaches the city centre without parking). It is shown that the population splits into two groups, agents who persistently choose parking spots close to the city centre on the one hand and agents settling for spots further away on the other. Grouping of agents in an economic context was studied in multi-resource minority games \cite{groupingmultiresourceMG}. In this model, grouping emerges when the probability that an agent will copy the strategy of a winning neighbour is large enough. However, in contrast to our model, it was assumed in this scenario that there is a considerable amount of structure in the connectivity among traders, as well as perfect information about the actions of neighbours.

\section{Model}
\label{Model}

We consider a simplified model of markets and decision-making traders with the aim of investigating the segregation of traders.
During each trading period agents are confronted with a choice of actions: where to trade -- \textit{choice of market} -- and how to trade -- \textit{whether to act as buyer or seller}. Decisions are made based on the attractions, which are accumulated scores an agent has received when taking actions in the past. The attractions to the various actions are updated after every trading period using a reinforcement learning rule of the form\footnote{Ref. \cite{luckystar} uses the same rule with $\omega=1-r$, while in Ref. \cite{CrutchfieldSato2003} the prescription used was $A(n+1)=S(n)+(1-\alpha)A(n)$. The second rule allows the attractions to increase to infinity, while in the first case, they are constrained. However, up to a temperature rescaling, the two rules are equivalent. The more important difference is that in the paper \cite{CrutchfieldSato2003}, the attractions of unplayed actions are updated with fictitious scores an agent would have got had he played the action, while we effectively update them 
with score $S(n)=0$.} 
\begin{align*}
A_{\gamma}(n+1) = \begin{cases}
(1-r)A_{\gamma}(n) + rS_{\gamma}(n), & \text{if agent has chosen action }\gamma\\
(1-r)A_{\gamma}(n), & \text{if agent has chosen action }\beta\neq\gamma 
\end{cases}
\label{reinfrule}
\end{align*}
where $S_{\gamma}(n)$ is the return gained by taking action $\gamma$ during $n$-th trade; $r$ is the parameter that describes the agent's memory. Its intuitive meaning is that each attraction is effectively an average of the returns over a shifting time window covering the previous $\frac{1}{r}$ trades. Finally, $A_{\gamma}(n+1)$ is attraction to the action $\gamma$ after $n$ trades, which will determine the action chosen in the following $(n+1)$-st trading period. 
The choice of action is then calculated using the \textit{softmax}\footnote{The softmax function is commonly used in models of learning agents, see for example \cite{luckystar}, \cite{CrutchfieldSato2003}. Another common formulation of the softmax function is $P_\gamma\propto\exp{(\beta A_\gamma)}$, where $\beta=1/T$ is sometimes called \textit{the intensity of choice} as in Ref. \cite{BrockHommes}.} function: the probability of taking an action $\gamma$ is $P_{\gamma} \propto \exp{(A_{\gamma}/T)}$.
The temperature $T$ regulates how strongly agents bias their preferences towards the option that gathered them the highest score. For $T\rightarrow0$ agents strictly choose the option with the highest attraction, while for $T\rightarrow\infty$ they choose randomly among the options.

\begin{figure}[h!]
 \vspace{-5pt}
        \centering
             \includegraphics[width=\textwidth]{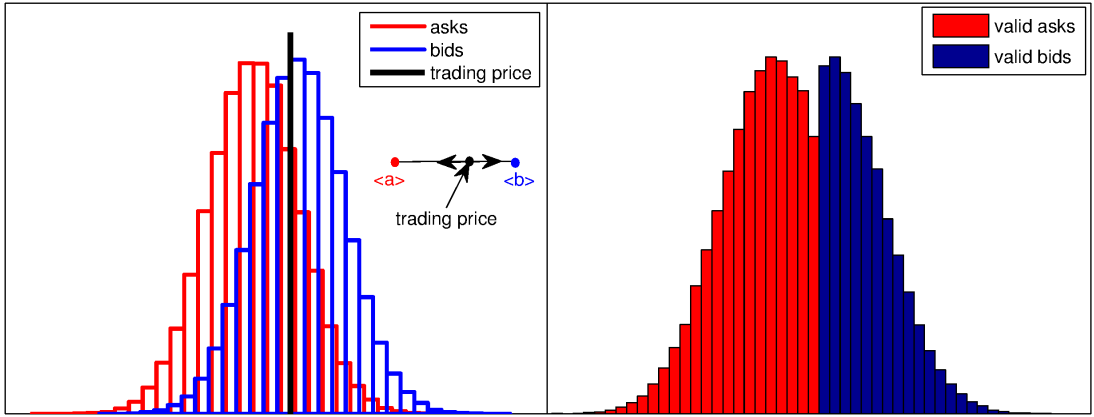}
                \caption{Illustration of market mechanism. (Left) Histogram of bids and asks arriving at a given market. The inset shows how the trading price is set, with a bias towards average bid or average ask regulated by the bias parameter $\theta$ of the market. (Right) Once invalid orders are eliminated, i.e.\ bids below or asks above the trading price, the distributions of valid bids and asks remain. Traders who have submitted valid orders are matched in random buyer-seller pairs for trading.}
        \label{marketmech}
\end{figure}

Orders to buy/sell at a certain price (bids and asks) are generated by traders independently of previous success or any other information; the bids and asks are independently identically distributed random variables (thus Zero Intelligence). We assumed bids ($b$) and asks ($a$) are normally distributed ($a \sim \mathcal{N}(\mu_a,\sigma_a^2)$ and $b \sim \mathcal{N}(\mu_b,\sigma_b^2)$), with means satisfying $\mu_b>\mu_a$. The assumption that the average bid is higher than the average ask is not crucial; it mainly allows a larger number of successful trades as the resulting trading price is typically below the average bid and above the average ask. In the work of Gode and Sunder (Ref.~\cite{ZItraders}) various demand and supply curves were used and thus both orderings of average bids and asks, $\langle a\rangle>\langle b\rangle$ and $\langle a\rangle<\langle b\rangle$, were investigated: they lead qualitatively to the same results. We similarly explored the case $\mu_a>\mu_b$, and apart from the obvious 
quantitative consequence that a smaller fraction of orders is valid for trade and consequently the number of successful trades is smaller, the qualitative results remain the same.
Once all traders have submitted an order to the market of their choice, then at each market the average bid $\langle b\rangle$ and average ask $\langle a\rangle$ are calculated and the trading price is set as $\pi=\langle a\rangle+\theta(\langle b\rangle-\langle a\rangle)$ with $\theta$ being a parameter that describes the bias or the market towards buyers (for $\theta<1/2$) or sellers (for $\theta>1/2$)\footnote{Note that traders are not informed about these market biases, nor the market mechanism in general; they learn only by means of the scores they receive.}.
All buyers who bid less and all sellers who ask more than the trading price are removed from the trading pool, as their orders cannot be satisfied at the price that has been set. The remaining traders are matched in random pairs of buyers and sellers, giving a total number of trades $\min\left(N_{\text{valid bids}}, N_{\text{valid asks}}\right)$.
For traders who manage to trade, the score is calculated as:\newline
\vspace{-10pt}
\begin{itemize}
\item[] $S(n)=\pi-a_n$ (sellers value getting more than they were asking for, i.e.\ $a_n$)
\item[] $S(n)=b_n-\pi$ (buyers value when they pay less then they intended, i.e.\ $b_n$)
\end{itemize}
All traders who do not get to trade receive return $S(n)=0$, and all orders are deleted from the market after each trading period. Figure \ref{marketmech} illustrates this market mechanism.

The assignment of returns that we are using was introduced in Ref.~\cite{ZItraders}, where it is associated with budget constraints of "Zero Intelligence-Constrained" traders. Exactly these agents were shown to reproduce the efficiency of human traders in double auction markets. In the original work, $a_n$ are cost values assigned to sellers, while $b_n$ are redemption values assigned to buyers. Traders were allowed to trade only if the trading price was lower than the redemption value or higher than the cost value, thus the name \textit{constrained} agents. Although the assignment of returns is the same in our model, we do not use the term budget constrained in the description as our agents are allowed to persistently buy (or sell), which is possible only if there is no overall wealth constraint\footnote{We note that also in Ref.~\cite{ZItraders}, agents were preassigned the role of a buyer or a seller and were not allowed to change this during trading, thus acting as if there was no overall constraint on 
the possession of money/goods for trade.}.
In our model the bids and asks could similarly be interpreted as cost and redemption values. We assume in addition that agents set orders based on these values, while the actual trading price is a function of the population averages.

\section{Results and Discussion}
\label{Results and Discussion}

In this section we will present the results from the simulations of the trading system described so far in this paper. 
Every system was defined in terms of the number of agents $N$, the number of markets $M=2$, the biases of the markets $\theta_1, \theta_2$,  the means and standard deviations of the distributions of bids and asks $\mu_a, \sigma_a, \mu_b, \sigma_b$, the temperature $T$ and the forgetting rate $r$. For every set of parameters simulations were run for $10,000$ trading periods; statistics are presented gathered from the last 100 trading periods of 100 independent runs of the stochastic dynamics.

In our system each agents has four preferences $p_{\mathcal{B}1}$, $p_{\mathcal{B}2}$, $p_{\mathcal{S}1}$, $p_{\mathcal{S}2}$ for the four possible actions of buying and selling at market 1 or 2. 
In the figures below, to help visualization we represent each agent by their total preference for buying ($p_{\mathcal{B}}=p_{\mathcal{B}1}+p_{\mathcal{B}2}$) and for market 1 ($p_1=p_{\mathcal{B}1}+p_{\mathcal{S}1}$). This is convenient as the corners in the $(p_\mathcal{B}, p_1)$ plane then represent the four \textit{pure strategies} -- agents \textit{always} buying at market 1, etc. Similarly, in the space of attractions we use two coordinates $(\Delta_{\mathcal{BS}}, \Delta_{12})$, which are basically attraction to buying as against selling and attraction to market 1 as against market 2. 

\begin{figure}[h!]
 \vspace{-5pt}
        \centering
			\subfigure[]{\includegraphics[width=0.49\textwidth]{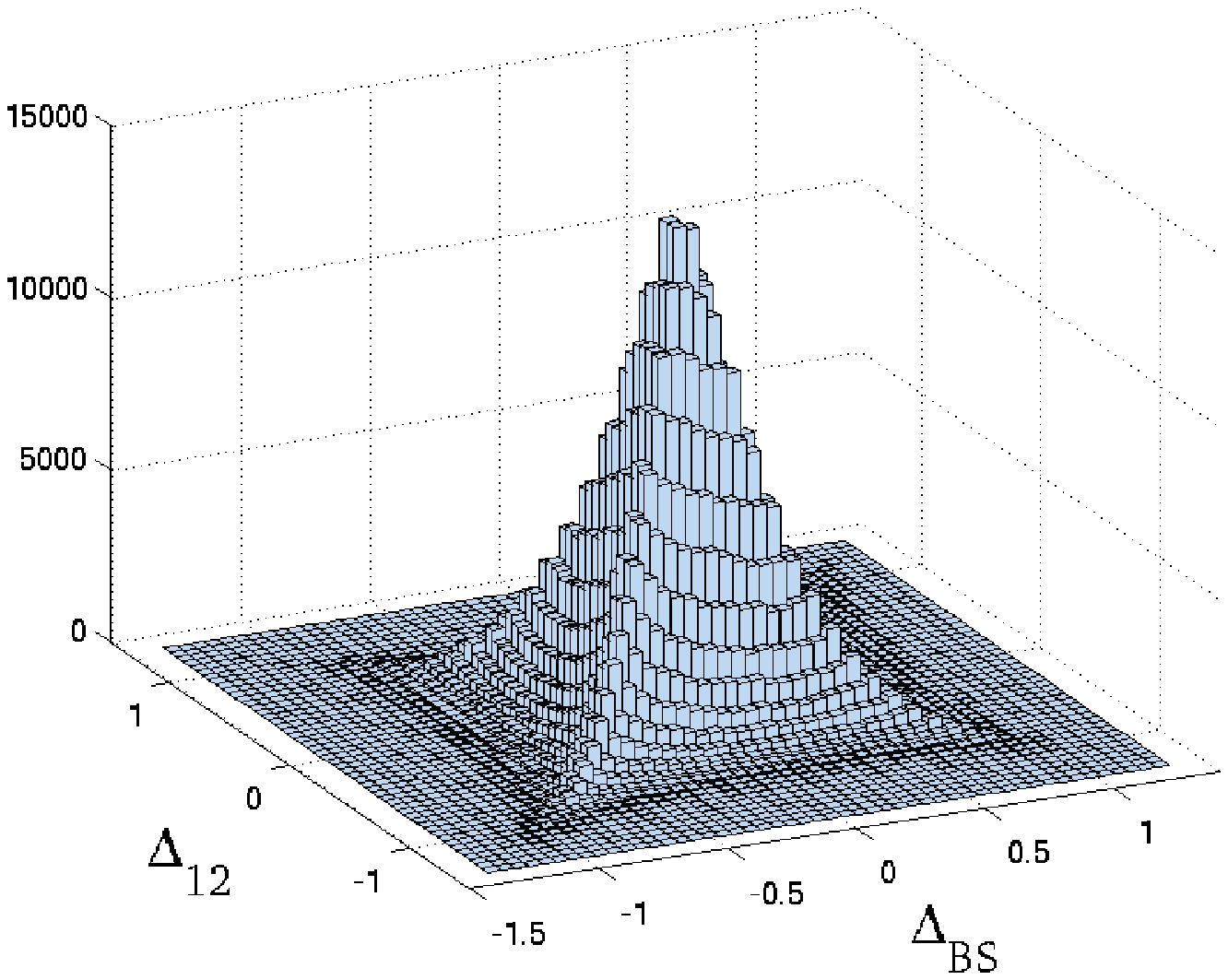}\label{AT029}}
			\subfigure[]{\includegraphics[width=0.49\textwidth]{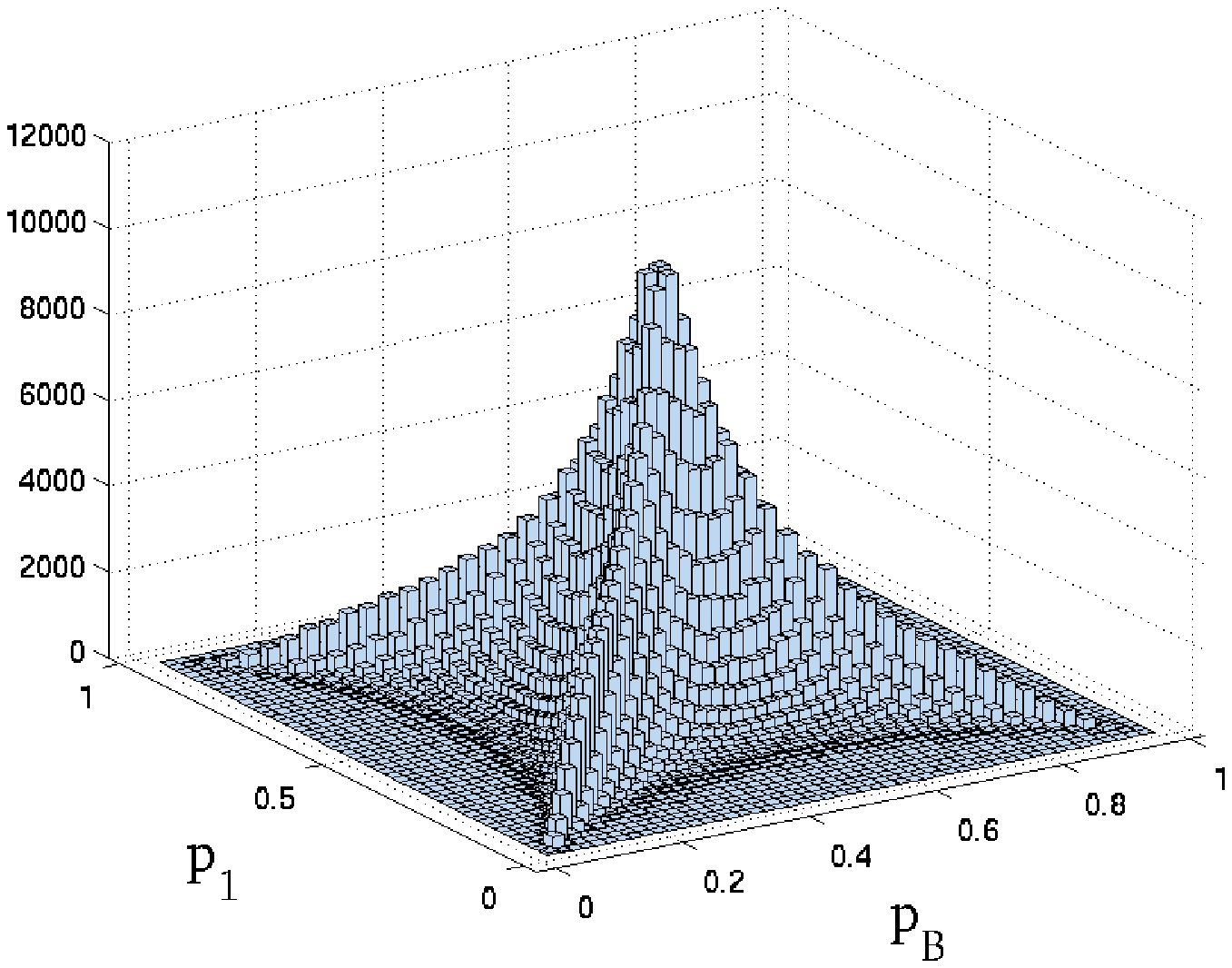}\label{PT029}}
        \caption{Steady state distributions at temperature $T=0.29$, with other parameters set to $N=200$, $M=2$, $\theta_1=0.3$, $\theta_2=0.7$, $r=0.1$, $\mu_b-\mu_a=1$, $\sigma_a=1$, $\sigma_b=1$. ~\ref{AT029}: Distribution of attractions.~\ref{PT029}: Distribution of preferences.}
         \vspace{-5pt}
        \label{unsegregated}
\end{figure}

In Figure \ref{unsegregated} we present steady state attraction and preference distributions for temperature $T=0.29$. An initially narrow, \textit{delta peaked} distribution (all scores are equal to 0) has been broadened due to diffusion arising from the random nature of returns. This steady state represents unsegregated behaviour of a population of traders. While the population does include some traders with moderately strong preferences for one of the actions, preferences remain weak on average. The population as a whole remains homogeneous in the sense that there is no split into discernible groups.
\begin{figure}[h!]
 \vspace{-5pt}
        \centering
             \subfigure[]{\includegraphics[width=0.49\textwidth]{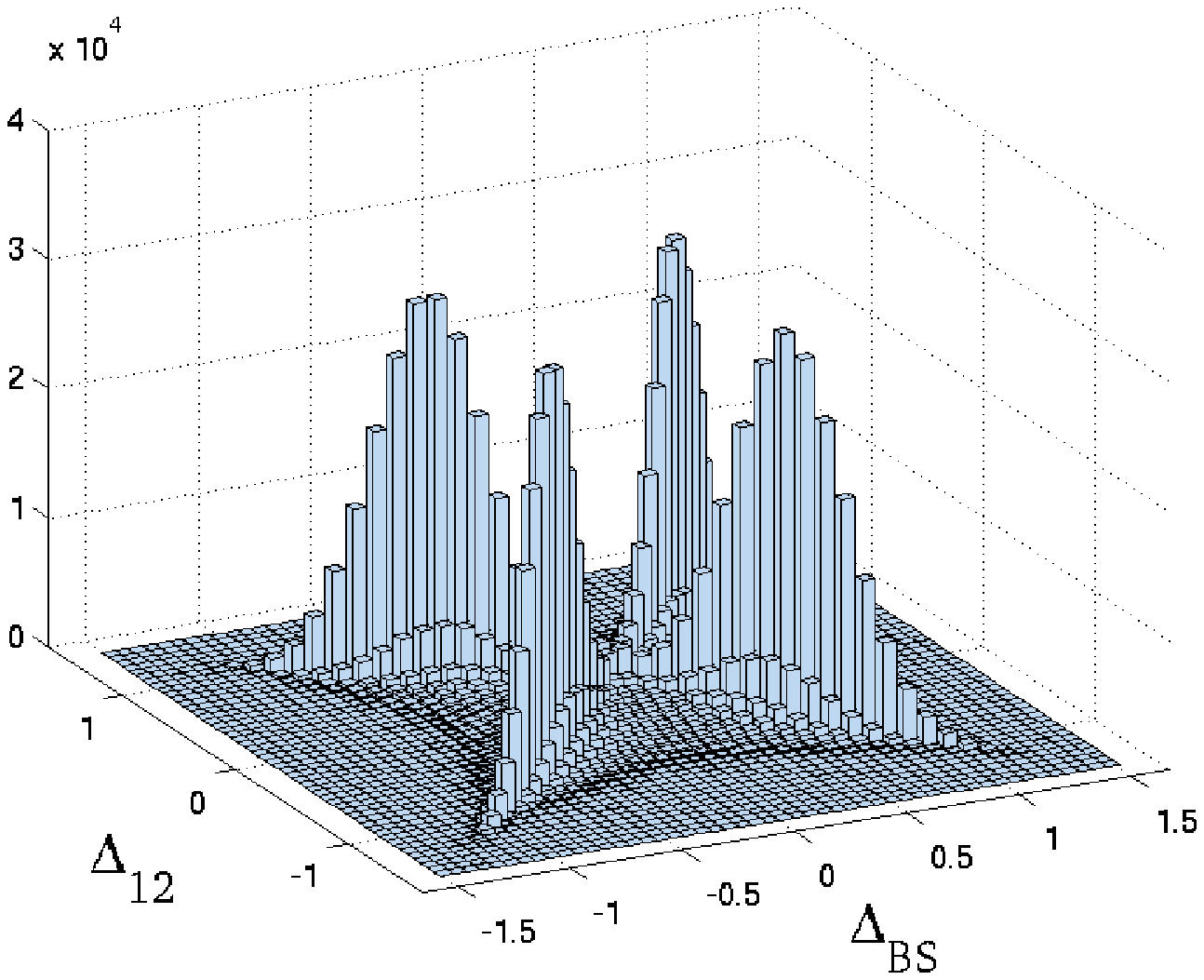}\label{AT014}}                
             \subfigure[]{\includegraphics[width=0.49\textwidth]{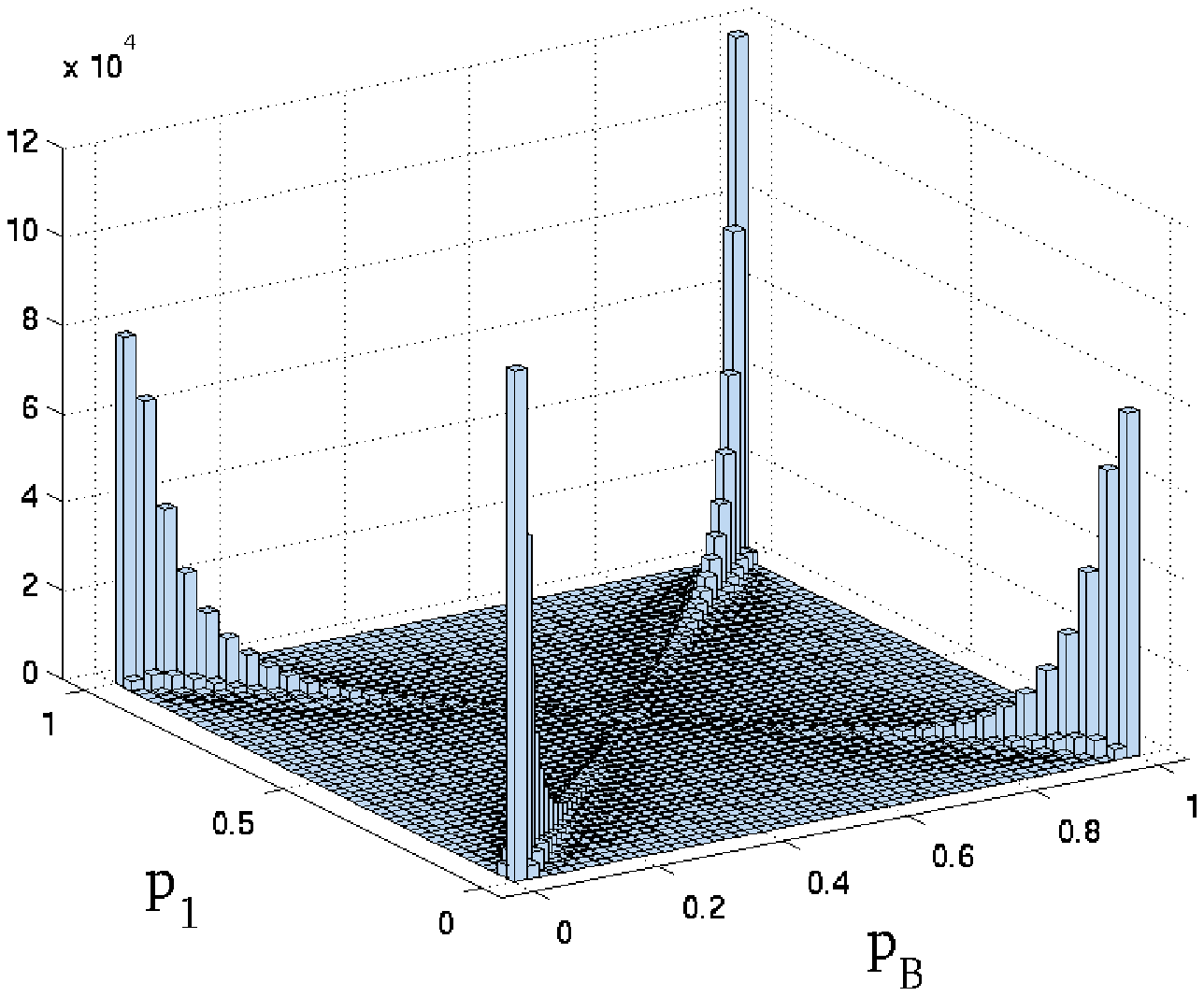}\label{PT014}}
        \caption{Steady state distributions in the low temperature regime ($T=0.14$, all other parameters as in Fig.~\ref{unsegregated}), showing clearly the segregation of traders into groups.~\ref{AT014}: Distribution of attractions.~\ref{PT014}: Distribution of preferences.}
        \vspace{-5pt}
        \label{segregatedsys}
\end{figure}

Figure \ref{segregatedsys} contrasts this scenario with the steady state of a system with exactly the same set of parameters but at the lower temperature $T=0.14$. The population of traders now splits into four groups, with the agents persistently trading at one of the markets, and thus we call this state \textit{segregated}. 
The markets shown in this example (Figs.~\ref{unsegregated}, \ref{segregatedsys}) are biased so that if an agent buys at market 1, or sells at market 2 (actions $\mathcal{B}1$ or $\mathcal{S}2$) he is awarded with a higher score. The traders who prefer these actions are \textit{``return-oriented traders''}. However, if all traders were return-oriented, they would have no partners for trading, and consequently they would received zero scores. To enable trading, some traders have developed strong preferences for buying (selling) at a market that gives them a lower average return ($\mathcal{B}2$, $\mathcal{S}1$). A larger fraction of these traders will be removed from the market as their orders will be regarded as invalid more frequently. Consequently, these traders will form a minority group and they will always find a trading partner, hence we will call them
\textit{``volume-oriented traders''}. The occurrence of segregation of an initially homogeneous population of traders into groups of return-oriented and volume-oriented traders is the main qualitative result of this paper. 

\begin{figure}[h!]
       \centering
             \includegraphics[width=0.7\textwidth]{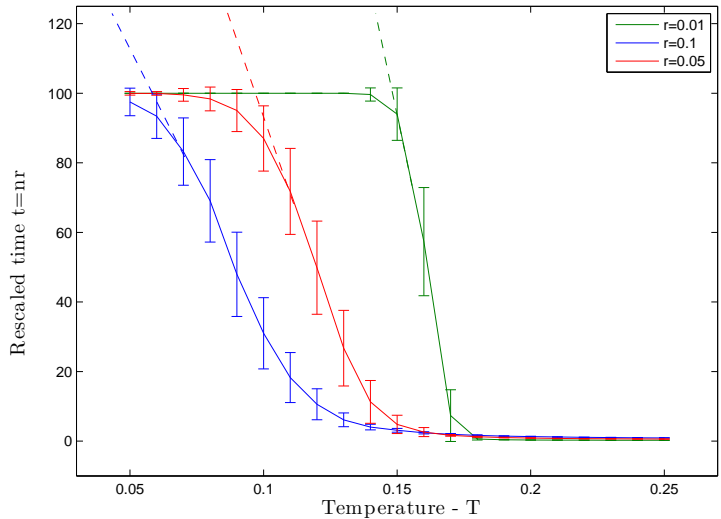}
                \caption{Average time an agent persists in any one of the four preference quadrants, plotted against temperature for different values of the forgetting rate, $r=0.1$ (blue), $r=0.05$ (red) and $r=0.01$ (green). Dashed lines are sketches of how the persistence times would increase further if they were not limited by the length of our simulation runs. Other parameters (as previously): $N=200$, $M=2$, $\theta_1=0.3$, $\theta_2=0.7$, $\mu_b-\mu_a=1$, $\sigma_a=1$, $\sigma_b=1$.}
        \label{times}
\end{figure}

When assessing stationarity of our system we measured population and time averages for various observables ($A_\gamma$, $\Delta_{\mathcal{BS}}$...). Depending on parameters, a stationary state was generally reached reasonably quickly, mostly within the first $1,000$ trading periods. Apart from stationarity we also investigated to what extent our system is ergodic, i.e.\ we wanted to exclude possibility that distributions in the low temperature regime might be a consequence of some agents' preferences becoming essentially \textit{frozen} after the first few trades. Quantitatively, we measured persistence times in one of four quadrants -- ``prefer buying at market 1'' ($\Delta_{\mathcal{BS}}>0$ and $\Delta_{12}>0$), ``prefer selling at market 1'' ($\Delta_{\mathcal{BS}}<0$ and $\Delta_{12}>0$), etc. Figure \ref{times} shows the average time an agent spent in any one of these quadrant before leaving it for another quadrant, for various temperatures. We present these plots for different values of the forgetting 
rate $r$, and using the rescaled time $t=rn$, where $n$ is the number of trading periods. (The use of $t$ rather than $n$ ensures that the trivial effect on persistence times of agents updating their attractions more slowly at smaller $r$ is removed.) From the Figure one sees that at small enough $r$, the onset of segregation is accompanied by a rapid increase in persistence times, showing that in the segregated state agents do indeed remain "loyal" to a given market for long times. On the other hand, we see that when temperatures are not too low (i.e.\ above the levelling off of the small-$r$ curves in Fig.~\ref{times}) then persistence times are short compared to the overall length of our runs, so that the system is ergodic.

\begin{figure}[h!]
       \centering
             \includegraphics[width=0.7\textwidth]{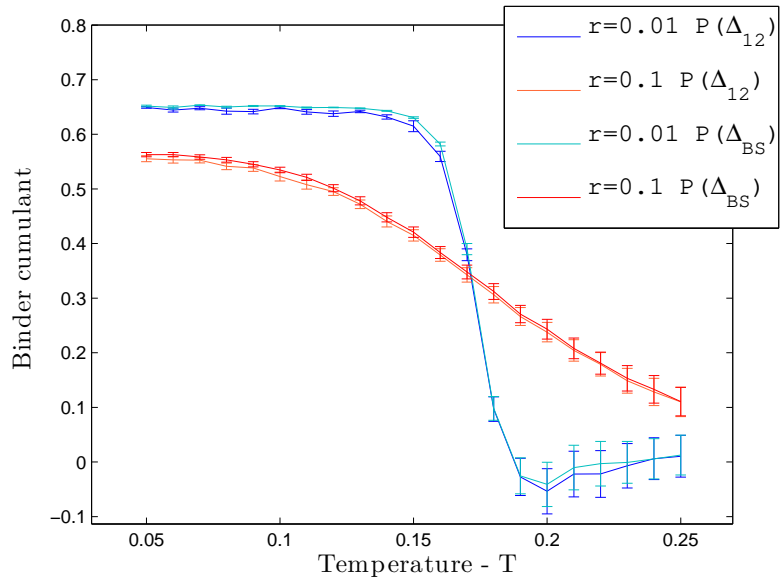}
                \caption{Binder cumulant for $P(\Delta_{BS})$ and $P(\Delta_{12})$ distributions, averaged over last 100 trading periods versus temperature for two different values of the forgetting rate, $r=0.1$ and $r=0.01$. Other parameters as previously $N=200$, $M=2$, $\theta_1=0.3$, $\theta_2=0.7$, $\mu_b-\mu_a=1$, $\sigma_a=1$, $\sigma_b=1$.}
        \label{binder}
\end{figure}

To quantify the observed change in the distributions of agent attractions or preferences as we go from unsegregated to segregated states, we measured higher cumulants of the distributions $P(\Delta_{\mathcal{BS}})$ and $P(\Delta_{12})$. Specially we tracked the \textit{Binder cumulant}:
$B=1-\frac{\langle \Delta^4\rangle}{3\langle \Delta^2\rangle^2}$. Figure \ref{binder} shows values of this Binder cumulant for various temperatures of the system, with all other parameters being same as in the previous figures. For higher temperatures, the Binder cumulant of our distributions approaches value characteristic of Gaussian distributions ($B=0$) as expected. At the other extreme, in the low temperature regime, the cumulant approaches a second characteristic value $B=2/3$, which is the Binder cumulant of a distribution consisting of two sharp peaks with equal weight. 
The transition between these two regimes is sharper for smaller values of $r$, making it possible to estimate the critical temperature for the onset of segregation.

Our simulation results suggest that even our simplified trading system shows rich and interesting behaviour. There exists a critical temperature $T_c$, such that for values $T<T_c$ the system segregates, i.e.\ the population of initially homogeneous traders splits into groups that persistently choose to trade at a specific market. The persistence times increase strongly with decreasing forgetting rate $r$ (see Fig.~\ref{times}) and we conjecture that in the limit $r\to0$ there is a sharp transition at $T_c$ in the sense that the persistence time diverges there. The exact value of the critical temperature is a function of the market parameters, and for the values of $\theta_{1,2}$ used above, we would estimate it from Figure \ref{binder} to be $T_c\approx 0.17$.

To understand in more detail how segregation arises, and the nature of the transition to the segregated state as $T$ is lowered, a simple mathematical description would evidently be useful. To obtain such a description, we can build on the approach of
Ref. \cite{BrockHommes}. This work studies the dynamics of agents who have to decide whether to purchase a sophisticated price predictor, or use a freely available naive predictor of price. This scenario differs from our model in a number of ways; apart from the more sophisticated trading strategies of the agents, it assumes perfect information about previous prices and about the performance of any price predictor. What is important in the analysis of Ref.~\cite{BrockHommes}, however, is that the limit of a large population of agents is implicitly taken, so that the system can be described entirely in terms of the fraction of agents choosing a given action (price predictor) at any instant in time, with these fractions evolving deterministically in time.
The authors of Ref.~\cite{BrockHommes} show that depending on the temperature, or the "intensity of choice" $\beta=1/T$, these two fractions can exhibit rich dynamics. The origin of this is that when all traders use sophisticated predictors, the cost of this predictor leads some agents to start choosing the free predictors, while there is a reverse effect from positive feedback when all traders use the simple predictor.

To adopt a similar approach for our model, we realize that mathematically our dynamics is Markovian, provided that we keep track of the attractions $A_\gamma^i$ to all actions $\gamma=\mathcal{B}1, \mathcal{S}1, \mathcal{B}2, \mathcal{S}2$ of all agents $i=1,\ldots,N$. Working with this description in a $4N$-dimensional continuous state space is, however, very difficult. As in Ref.~\cite{BrockHommes} we can therefore consider the large $N$-limit where the trading price at each market is no longer affected by fluctuations in the number and value of orders submitted. We also consider the limit of small $r$, using as time unit again the rescaled time $t=rn$ so that a unit time interval in $t$ corresponds to $1/r$ trading periods.
The fluctuations in each individual agent's attractions then also tend to zero because they are averaged over many ($\sim 1/r$) returns each contributing a small ($\sim r$) change of attraction. As long as the agent population remains homogeneous, all agents should in the limit have the same attractions $A_\gamma$. In that case the system is described entirely in terms of the average values of these four attractions, or correspondingly the fraction of agents choosing each of the four options $\gamma$. As these fractions add to unity it is enough to keep track of three of them, and one can write down deterministic equations for their time evolution.(Details are beyond the scope of this paper and will be given elsewhere.)

\begin{figure}[h!]
 \vspace{-5pt}
        \centering
        \subfigure[$T=0.32$]{\includegraphics[width=0.46\textwidth]{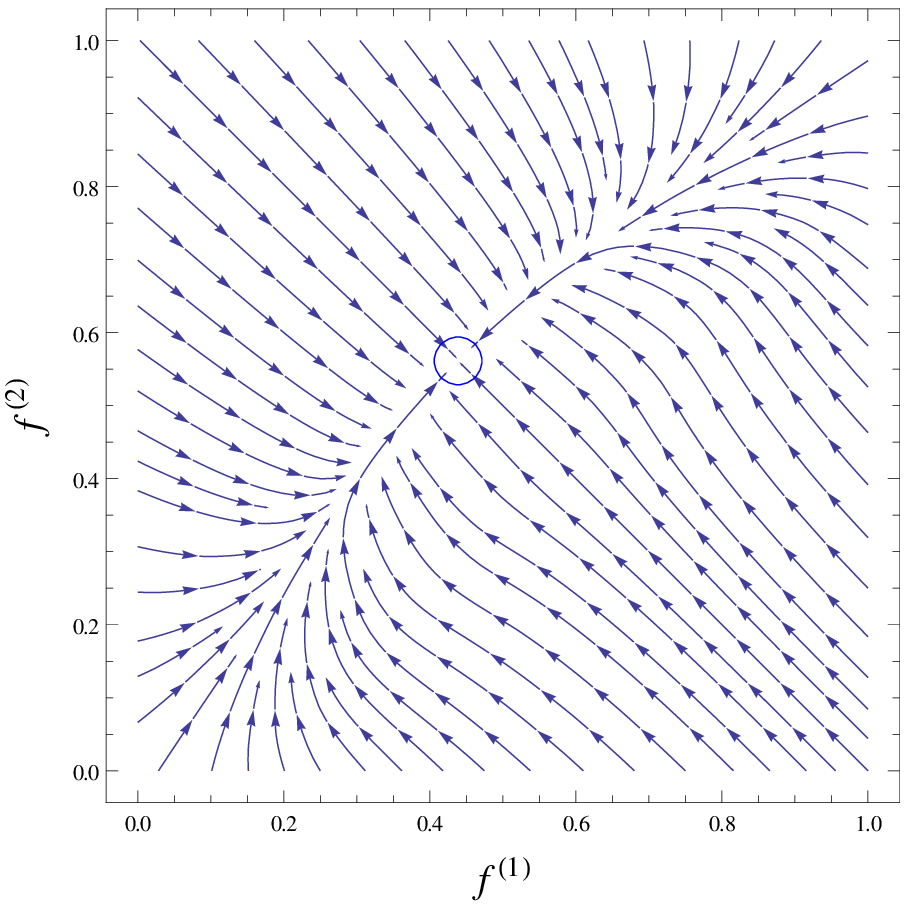}\label{highT}}
        \subfigure[$T=0.29$]{\includegraphics[width=0.46\textwidth]{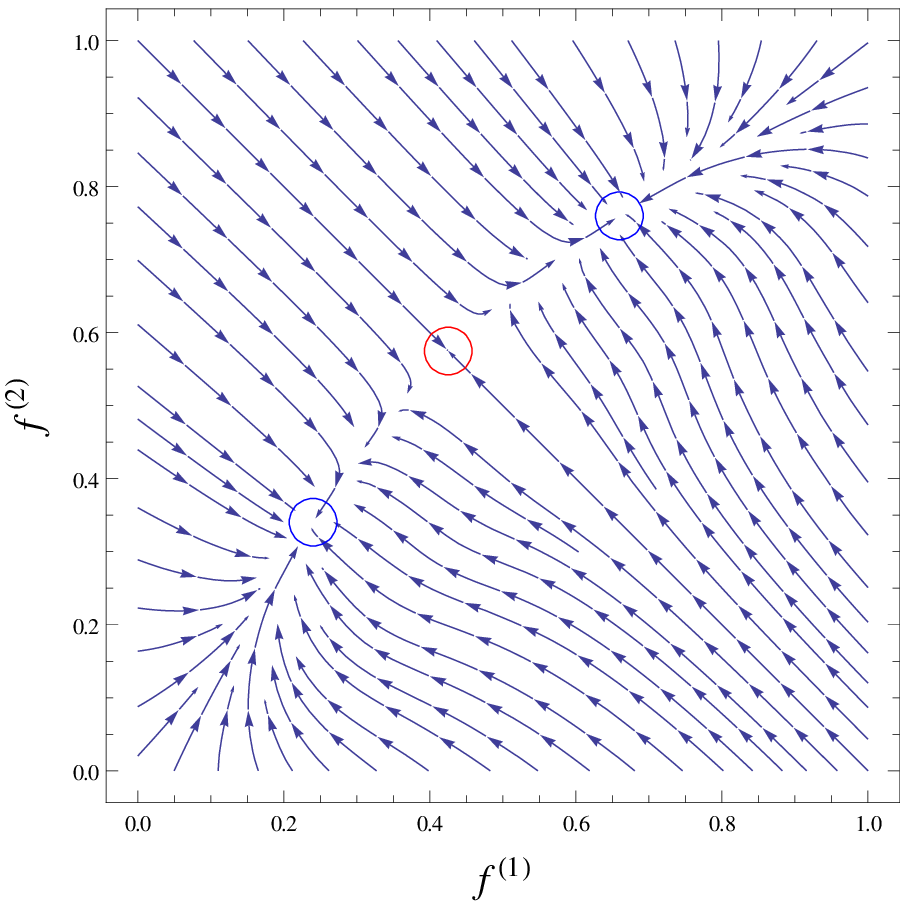}\label{lowT}}
           \caption{Flow diagrams that describe the large population dynamics of our two-group model in the space of fractions of agents from each group who choose market 1. $f^{(1)}$ is the fraction going to market 1 in the group of agents who typically sell ($p^{(1)}_{\mathcal{B}}=0.2$), and $f^{(2)}$ the corresponding fraction in the group of ``buyers'' ($p^{(2)}_{\mathcal{B}}=0.8$). The markets have symmetric biases $\theta_1=0.3=1-\theta_2$. ~\ref{highT} High temperature: the dynamics has a single fixed point. ~\ref{lowT} Low temperature: single fixed point has become unstable.}
        \vspace{-5pt}
        \label{analyticalres}
\end{figure}

The results of the above approach for our model are still somewhat difficult to visualize as we need to track fixed points and trajectories in a three-dimensional space. We therefore switch to a simpler system that gives qualitatively similar results: a population of traders consisting of two equal-sized groups with fixed preference for buying $p^{(1)}_{\mathcal{B}}$ and $p^{(2)}_{\mathcal{B}}$, respectively. The agents then only choose between two actions, namely, whether to go to market 1 or 2 in each trading period. Although the system where agents change their buy-sell preferences is more plausible behaviourally, the two-group model still undergoes segregation and requires us (for $N\to\infty$, $r\to 0$ and assuming an unsegregated state as above) to track only the fraction of agents choosing market 1 in each of the two groups. We denote these fractions by $f^{(1)}$ and $f^{(2)}$.
In Figure \ref{analyticalres} we present the flow diagrams that we find for the time evolution of these two fractions, at high and low $T$. At high temperature, one observes a single fixed point as expected (Fig~\ref{analyticalres} (a)). As $T$ is lowered, this fixed point becomes unstable, and two additional stable fixed points appear ((Fig~\ref{analyticalres} (b)). The temperature where the high-$T$ fixed point first becomes unstable thus identifies the critical temperature $T_c$ for the onset of segregation. We also find that the new stable fixed points evolve continuously from the high-$T$ fixed point as $T$ is lowered through $T_c$, so the segregation transition has the character of a bifurcation and is continuous.

It is worth emphasizing that the locations of the new fixed points that appear at low temperature are not necessarily meaningful: as explained above, the simplifications that have allowed us to consider deterministic time evolution in a simple two-dimensional space require that the agent population remains homogeneous. By construction, this simple picture can therefore not describe quantitatively the segregated populations of agents that arise below $T_c$. Nevertheless, the instability of the high-$T$ unsegregated fixed point is enough to identify the temperature for the onset of segregation.

\begin{figure}[h!]
             \includegraphics[width=0.6\textwidth]{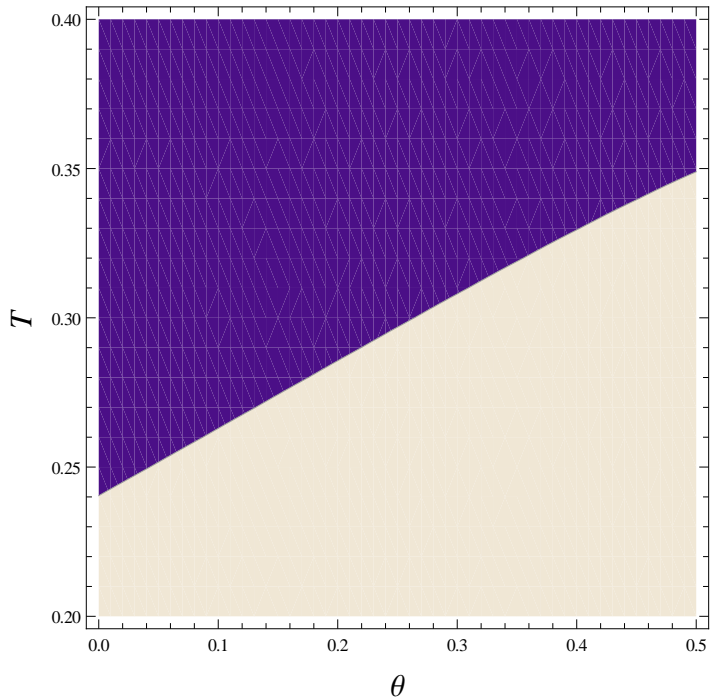}
                \caption{Segregation temperature $T_c$ versus market bias $\theta_1=1-\theta_2=\theta$. In this diagram $T_c(\theta)$ separates segregated (light blue) and unsegregated (dark blue) steady states. Results are shown for the two-group model with the two groups of agents having fixed buying preferences of 
                 $p^{(1)}_{\mathcal{B}}=0.2$ and $p^{(2)}_{\mathcal{B}}=0.8$, respectively.}
        \label{phasediag}
\end{figure}

The analytical description sketched briefly above allows us to study, for example, how the value of the critical temperature $T_c$ depends on the parameters of the problem, specifically for the two-group model on $p^{(1)}_{\mathcal{B}}$ and $p^{(2)}_{\mathcal{B}}$ and on the market biases $\theta_1$ and $\theta_2$. As an example, Figure \ref{phasediag} shows how $T_c$ varies with the market bias, still for the case of symmetric markets $\theta_1=1-\theta_2=\theta$. One sees that for every value $\theta$ there exists a critical temperature $T_c$ at which a bifurcation to a segregated steady state occurs. Note that the temperature region where segregation occurs {\em shrinks} as the difference between the market biases {\em increases} (smaller $\theta$), showing that segregation is a collective effect rather than being trivially driven by the differences between the markets.
For $\theta=0.3$ as in Fig.~\ref{analyticalres}, one finds $T_c(\theta)=0.308$. From simulations for a system with $N=100$ traders and forgetting rate $r=0.1$, we estimate a value of $T_c\approx0.3$. This is an excellent agreement with the theoretical prediction, especially bearing in mind that the latter applies directly only to the limit $N\to\infty$ and $r\to 0$. 

In our original model where the agents can adapt their preferences both for the two markets and for whether to buy or sell, the quantitative agreement is slightly less good. E.g.\ for $\theta_1=1-\theta_2=0.3$ and bid and ask distributions parameters as in Figs.~\ref{unsegregated},~\ref{segregatedsys} the analytical description predicts $T_c\approx 0.127$.  Our simulations for a population of $N=200$ traders with forgetting rate $r=0.1$, on the other hand, lead to the estimate $T_c\approx 0.2$. This suggests that in the fully adaptive model the effects of nonzero forgetting rate and finite population size are stronger than in the two-groups model.

\section{Concluding remarks}
\label{Concluding remarks}

With so much trade and commerce moving online over the last two decades, the study, design, operation, and good governance of electronic marketplaces has become a major area of computer science, both theoretical and applied.    Much online economic activity --- for example, most trading in western financial markets --- is  now undertaken by automated computer programs, which are software agents acting on behalf of human principals or companies.  A key research goal in the study of electronic marketplaces is, therefore, to understand the long-run dynamics of these markets when populated by automated software traders.  This leads to questions such as:  what long-run states are possible in these marketplaces, what patterns in states occur or recur, what states may be avoided and how, what states may be encouraged to occur and how, etc.      The practical economic and financial consequences of such understanding are immense.  The so-called \emph{Flash Crash} of US stock markets on 6 May 2010 showed the 
vulnerability of inter-linked trading systems to a single large trade, for example, and has led to the implementation of automated “circuit breakers” to eliminate or reduce the sector-wide impacts of rapid market movements \cite{govref}.   The importance of these issues is shown by the establishment of a major research programme by the UK Government’s Department of Business, Industry and Skills on computer trading in financial markets%
\footnote{See:  \url{http://www.bis.gov.uk/foresight/our-work/projects/published-projects/computer-trading}}. Our research in this same vein focuses on a description of a specific characteristic of trading systems --- segregation. As argued in the introduction, specialized (segregated) traders might be better in terms of exploitation of a market. However with specialization there comes an associated vulnerability as agents become more exposed to losses if all their investments are focused on a single market that might crash. Ultimately, we would like to describe and predict the long-run dynamics of marketplaces comprising automated interacting traders and to extract a set of regulations that might promote or suppress the segregation.

In this paper we introduced a simplified model of double auction mechanisms with Zero Intelligence traders, with the goal of investigating the possibility of a spontaneous segregation of traders. The use of ZI traders was motivated by the hypothesis that segregation can emerge as a consequence of market mechanisms and learning rules, neglecting complexity in trading strategies. We presented results form numerical simulations and outlined how analytical methods can be used to understand the occurrence of segregation, giving quantitatively reasonable predictions even away from the limits (infinite population of traders, infinitesimal forgetting rate) where the analysis is derived.
Although the relevance of our model with respect to real economies might be questioned due to its simplicity, it is interesting to note that even in this simple trading mechanism, learning agents who interact only via markets can end up being segregated.

\end{document}